\documentclass[aps,prb,twocolumn,groupedaddress,showpacs,10pt]{revtex4-1}
\usepackage{amsmath}
\usepackage{amssymb}
\usepackage{bm}
\usepackage[utf8]{inputenc}
\usepackage{mathptmx}
\usepackage{graphicx}
\usepackage{multirow}
\usepackage{lmodern}
\usepackage{color}
\usepackage{units}
\usepackage{dcolumn}
\usepackage{comment}
\usepackage[bookmarks=true,colorlinks=true,linkcolor=blue,citecolor=blue,urlcolor=blue]{hyperref}
\usepackage[figure]{hypcap}

\begin{document}

\title{Relativistic first principles theory of Yu--Shiba--Rusinov states applied to an Mn adatom and Mn dimers on Nb(110)}
\author{Bendegúz Ny\'ari$^1$} 

\author{András Lászl\'offy$^{1,2}$}

\author{László Szunyogh$^{1,3}$}

\author{Gábor Csire$^{2,4}$}

\author{Kyungwha Park$^5$}

\author{Balázs Ujfalussy$^{1,2}$}
\affiliation{$^1$Department of Theoretical Physics, Budapest University of Technology and Economics, Budafoki \'ut 8., HU-1111 Budapest, Hungary}

\affiliation{$^2$Wigner Research Centre for Physics, Institute for Solid State Physics and Optics, H-1525 Budapest, Hungary}

\affiliation{$^3$MTA-BME Condensed Matter Research Group, Budapest University of Technology and Economics, Budafoki \'ut 8., HU-1111 Budapest, Hungary}

\affiliation{$^4$Catalan Institute of Nanoscience and Nanotechnology (ICN2), CSIC, BIST, Campus UAB, Bellaterra, Barcelona, 08193, Spain}

\affiliation{$^5$Department of Physics, Virginia Tech, Blacksburg, Virginia 24061, USA}

\date{\today}

\begin{abstract}
We present a  fully relativistic first principles based theoretical approach for the calculation of the spectral properties of magnetic impurities on the surface of a superconducting substrate, providing a material specific framework for the investigation of the Yu--Shiba--Rusinov (YSR) states. By using a suitable orbital decomposition of the local densities of states we discuss in great details the formation of the YSR states for an Mn adatom and for two kinds of Mn dimers placed on the Nb(110) surface and compare our results to recent experimental findings. In case of the adatom we find that the spin-orbit coupling slightly shifts some of the YSR peaks and also the local spin-polarization on the Nb atoms have marginal effects to the their positions. Moreover, by scaling the exchange field on the Mn site we could explain the lack of the $d_{x^2-y^2}$-like YSR state in the spectrum. While our results for a close packed ferromagnetic dimer are in satisfactory agreement with the experimentally observed splitting of the YSR states,  in case of an antiferromagnetic dimer we find that the spin-orbit coupling is not sufficiently large to explain the splitting of the YSR states seen in the experiment. Changing the relative orientation of the magnetic moments in this dimer induces splitting of the YSR states and also shifts their energy, leading even to the formation of a zero bias peak in case of the deepest YSR state.
\end{abstract}

\maketitle

\section{Introduction}\label{intro}

Local magnetic moments inside or at surfaces of metals in the superconducting state lead to the formation of localized bound states within the superconducting gap, referred to as Yu-Shiba-Rusinov (YSR) states \cite{1,2,3}.
In the simplest picture, the magnetic moment is viewed as a classical spin which is exchange-coupled to the Cooper pairs, and a single particle-hole symmetric pair of YSR states is predicted within the gap \cite{3,4,5,6}. With the high resolution of scanning tunneling microscopy (STM) and spectroscopy (STS) techniques available nowadays, experiments are able to resolve several pairs of YSR resonances within the superconducting gap \cite{9,10,11}. This has been explained due to the crystal field of the substrate which lifts the degeneracy of the adatom's $d$ level and, consequently, also the degeneracy of the YSR states \cite{6,9}.

While these studies revealed the mechanism of the formation of the YSR states in the magnetic impurity-superconductor systems in many details, a materials specific description is still missing. In the present paper, we adopt and further develop the solution of the Bogoliubov--de Gennes (BdG) equations with band theoretical methods treating, therefore, magnetism, relativistic (spin-orbit coupling) effects, superconductivity and the self-consistent electronic structure of the host and the impurity on the same footing \cite{100,101}. Such a solution incorporates the full orbital character of the electronic structure of real impurities
and is not only able to address both the general and the specific properties of multi-orbital YSR states, but also
allows a straightforward comparison to local experimental probes having direct access to the local density of states (LDOS).

Understanding inhomogeneous superconductors at the {\em ab initio } level is rather challenging, even for conventional electron-phonon coupled systems. Previously, some of the present authors demonstrated that by solving the Dirac-Bogoliubov--de Gennes (DBdG) equations via Multiple Scattering Theory (MST) is a very powerful tool to explain  induced superconductivity\cite{101,104}, induced triplet pairing in superconductors\cite{csire-relativistic2018} or the effects of Fermi surface anisotropy \cite{107}. Almost twenty years ago, the Embedded Cluster Method (ECM) has been employed with the MST to describe finite magnetic clusters of atoms in the normal state \cite{lazarovits-2002}. Recently, the non-relativistic BdG-MST and the ECM have been combined and used to study bulk impurity-superconductor systems \cite{saunderson-2021}.

In this work we extend this method by merging the fully relativistic DBdG-MST equations with the ECM to perform a first principles study of the YSR states induced by magnetic impurities at superconducting surfaces.
In the next section we briefly describe the elements of the theory and our computational implementation. Then we apply the newly developed code to an Mn adatom and Mn dimers on top of the (110) surface of bcc Niobium and we compare or results with the experimental observations\cite{beck-natcomm2021}. By decomposing the local density of states (LDOS) according to atomic orbitals we perform a thorough analysis of the YSR states and also investigate the effect of the spin-orbit coupling, of the local spin-polarization in the substrate and of the strength of the exchange field at the Mn site. We also consider two kinds of close packed Mn dimers, one with ferromagnetic (FM) and the other with antiferromagnetic (AFM) coupling. We analyze the hybridization and corresponding splitting of the YSR states in the FM coupled dimer. Our calculations show that this effect is strongly suppressed in the AFM dimer even in the presence of SOC.  Finally, we studied the effect of the noncollinearity of the spin moments in the AFM dimer and, in case of the deepest YSR state, we show the formation of a zero bias peak (ZBP) at a certain tilting angle. 

\section{Methods}\label{methods}
\subsection{The Dirac-Bogoliubov--de Gennes equation}\label{sec:bdg-skkr}
Following Ref. \onlinecite{csire-relativistic2018} we seek for the solutions of the
 Dirac--Bogoliubov--de Gennes (DBdG) equation,
\begin{equation}
 \left[ \varepsilon - \left(
 \begin{array}{cc}
H_{\text{D}} & \Delta_{\text{eff}}\eta \\
\Delta^\ast_{\text{eff}}\eta^T &  -H^\ast_{\text{D}} \\
\end{array}  \right) \right]
\left( \begin{array}{c} \Psi_e (\varepsilon)\\ \Psi_h (\varepsilon) \end{array} \right) = 0 \;, \label{bdg}
\end{equation}
where $\varepsilon$ is the energy relative to the Fermi energy $E_F$, $\Psi_e (\varepsilon)$ and  $\Psi_h (\varepsilon)$ are the four-component electron and hole part of the wavefunction, respectively, $\Delta_{\rm eff}$ is the effective pair interaction, while
\begin{equation}
H_{D}=c\vec{\alpha}\vec{p}+(\beta-\mathbb{I}_4)c^2/2+\left( V_\text{eff} - E_F \right)
( \vec{r})\mathbb{I}_4+\vec{\Sigma}\vec{B}_\text{eff}(\vec{r}) \, ,
\label{eq:Dirac}
\end{equation}
is the Kohn--Sham--Dirac Hamiltonian in Rydberg units ($\hbar=1$, $m=1/2$, $e^2=2$, $c=274.072$) with the effective atomic potential $V_\text{eff}(\vec{r})$ and effective exchange field $\vec{B}_\text{eff}(\vec{r})$. The $4 \times 4$ matrices in Eqs. \eqref{bdg} and \eqref{eq:Dirac} are defined as
\begin{equation*}
\vec{\alpha}=\left(
\begin{array}{cc}
0 & \vec{\sigma} \\
\vec{\sigma} & 0
\end{array}
\right) \, ,
\quad
\beta=\left(
\begin{array}{cc}
\mathbb{I}_2 & 0 \\
0 & -\mathbb{I}_2
\end{array}
\right) \, ,
\end{equation*}
\begin{equation*}
\vec{\Sigma}=\left(
\begin{array}{cc}
\vec{\sigma} & 0 \\
0 & \vec{\sigma}
\end{array}
\right)\, ,
\quad
\eta=\left( \begin{array}{cc}
i \sigma_y & 0  \\
0 & i \sigma_y
\end{array}
\right) \, ,
\label{eq:Bdg}
\end{equation*}
where $\vec{\sigma}$ stand for the Pauli matrices.
As described by for example in Ref. \onlinecite{101}, first
$V_\text{eff}(r)$ and $\vec{B}_\text{eff}(r)$ are calculated self-consistently in the normal state, then the solution of \eqref{bdg} proceeds by either fitting the $\Delta_{\rm eff}$ parameter to the experimental value of the energy gap in the bulk superconductor, or, by assuming a suitable approximate energy functional and completing the self-consistency cycle in the superconducting state \cite{107}. In the present paper we choose the simpler first route by fitting $\Delta_{\rm eff}$ to the Nb bulk superconducting gap. This seems to be a reasonable approximation, because, on the one hand, a magnetic adatom is not expected to support any singlet pairing interaction, therefore $\Delta_{\rm eff}=0$ can be assumed on the impurity sites,
while, on the other hand, the pairing in the superconducting host is not going to be influenced by a few impurity atoms.

\subsection{Embedded Cluster Green's function technique}\label{sec:cluster}

In order to describe an impurity on a superconducting surface, it is desirable to avoid the supercell approximation used frequently in normal-state calculations.
Namely, the superconducting coherence length is usually very large compared to the lattice constant, thus it is expected that
the interference between supercells  might seriously affect the obtained results.
The Screened Korringa-Kohn-Rostoker (SKKR) method
used to solve the DBdG equations for layered systems in Ref. \onlinecite{csire-relativistic2018} provides an excellent basis to achieve this goal.
First, the Green's function of a semi-infinite surface is calculated. The second step is the description of the surface impurity via the Green's function embedding technique \cite{lazarovits-2002}.
This procedure trivially leads to the Green's function of the impurity-superconductor system,
\begin{align}
&G^{\rm ab}(\varepsilon,\vec{r}+\vec{R}_i,\vec{r}^{\;\prime}+\vec{R}_j)=\sum_{QQ^\prime}Z^{{\rm a},R}_{i,Q}(\varepsilon,\vec{r}) \tau^{{\rm ab},ij}_{QQ^\prime}(\varepsilon) Z^{{\rm b},L}_{j,Q^\prime}(\varepsilon,\vec{r}^{\;\prime})\nonumber\\
&-\delta_{ij} \left\{ \theta(r^\prime-r)\sum_Q Z^{{\rm a},R}_{i,Q}(\varepsilon,\vec{r})J_{i,Q}^{{\rm b},L}(\varepsilon,\vec{r}^{\;\prime}) \right. \nonumber\\
& \quad \quad +\left. \theta(r-r^\prime)\sum_Q J^{{\rm a},R}_{i,Q}(\varepsilon,\vec{r})Z_{i,Q}^{{\rm b},L}(\varepsilon,\vec{r}^{\;\prime}) \right\}
\label{eq:gf}
\end{align}
where ${\rm a,b} \in \left\{ {\rm e}, {\rm h} \right\}$, $Q$ and $Q^\prime$ are total angular momentum indices, $i$ and $j$ denote sites with position vectors $\vec{R}_i$ and $\vec{R}_j$, respectively,
while $Z^{{\rm a},R/L}_{i,Q}$ and $J^{{\rm a},R/L}_{i,Q}$ are in order properly normalized regular and irregular right-hand-side(R) and left-hand-side(L) solutions of the DBdG equation \eqref{bdg} \cite{csire-relativistic2018}. In this equation the scattering due to the impurity  is taken into account by replacing the inverse scattering matrices ${\bm t}_\text{host}^{-1}(\varepsilon)$  of the host atoms by those of the impurity or a of a cluster of impurities ${\bm t}_\text{clus}^{-1}(\varepsilon) $ to obtain the matrix of the scattering path operator (SPO) at the impurity sites,
\begin{equation}
{\bm\tau}_{\text{clus}}(\varepsilon) = {\bm \tau}_\text{host}(\varepsilon)\left\{{\bm I}-\left[{\bm t}_\text{host}^{-1}(\varepsilon)-{\bm t}_\text{clus}^{-1}(\varepsilon) \right]{\bm \tau}_\text{host}(\varepsilon) \right\}^{-1}.\;
\label{eq:embedding}
\end{equation}
Here the bold symbols denote matrices with both angular momentum and site indices, whereas the latter ones are restricted to the sites of the impurity cluster.
For the details of the embedded cluster method see for example Ref. \onlinecite{lazarovits-2002}. Note that the solution of the DBdG equation instead of the Dirac equation implies a doubled matrix dimension for the SPO matrix
due to the appearance of the electron-hole index,
\begin{equation}
{\bm \tau}(\varepsilon)=
\left(
\begin{array}{cc}
{\bm \tau}^{\rm ee}(\varepsilon) & {\bm \tau}^{\rm eh}(\varepsilon) \\
{\bm \tau}^{\rm he}(\varepsilon) & {\bm \tau}^{\rm hh}(\varepsilon)
\end{array}
\right)\;.
\end{equation}
Once the SPO for the impurity system is obtained, the Green's function can be calculated according to Eq. \eqref{eq:gf}. It should be noted that the embedding process via Eq. \eqref{eq:embedding} requires the calculation of the site off-diagonal elements of the SPO. This is the most time consuming step of the calculation, however, it enables the embedding of a significantly larger cluster of impurities as compared to the ordinary size of the supercells in supercell methods. Once the Green's function is calculated, the local density of states (LDOS) averaged inside the atomic cell $V_i$ can be obtained as usual, albeit now including new component indices for the electrons and the holes,
\begin{equation}
n_{i}^{\rm ab}(\varepsilon)= -\dfrac{1}{\pi} \text{Im}  \int_{V_i} d^3r \, \text{Tr} \,
G^{\rm ab}(\varepsilon,\vec{r}+\vec{R}_i,\vec{r}+\vec{R}_i)
\;,
\label{eq:dos}
\end{equation}
where
the Tr denotes the trace in the four-component Hilbert space. In this work we will present electron and hole LDOS's, $n_{i}^{\rm ee}(\varepsilon)$ and $n_{i}^{\rm hh}(\varepsilon)$, respectively, and the sum of them termed as the total LDOS.

\section{Results}\label{sec:results}
In the present work we apply the theory described above to an Mn adatom and various Mn dimers on the Nb(110) surface, studied in detail recently by STM experiments and also theoretically using a tight-binding model \cite{beck-natcomm2021}.
The self-consistent calculations in the normal state were performed
by employing the local density approximation according to Vosko \textit{et al.} \cite{vosko-1980} and the atomic sphere approximation (ASA) for the description of the single-site effective potentials and exchange fields, while the scattering of the electrons was treated using an orbital momentum cutoff of $\ell_\text{max}=2$. For the energy integration we used a semi-circular energy contour on the upper half plane with 16 energy points. For the self-consistent calculation of the Nb(110) surface layers, forming two-dimensional (2D) centered rectangular lattices with $C_{2v}$ point-group symmetry,  we sampled 253 $k$ points in the irreducible wedge of the 2D Brillouin zone (BZ).  By neglecting geometrical relaxions, both in the absence and the presence of impurities, the lattice positions were considered in an ideal parent bulk geometry with a lattice constant of $a=6.237 \, a_0$ \cite{straumanis-1970}. The monomer and the dimer atoms were then placed in hollow positions above the surface layer of Nb(110), i.e. in the lattice positions of the first vacuum (empty sphere) layer.

To take into account charge- and spin-density oscillations caused by the impurity in the substrate, beyond the magnetic impurities we inlcuded substrate atoms and also empty spheres representing the vacuum in the self-consistently treated embedded cluster within a radius of $r=1.21 \, a_{\rm 2D}$ ($a_{\rm 2D}=\sqrt{2}a$), often referred to as an extended Impurity Cluster (IC).
By solving the DBdG equations for the extended impurity we calculated the LDOS for each embedded atom according to Eq. \eqref{eq:dos}. In this final step we increased the size of the IC to $3a_{\rm 2D}$ (including e.g. 118 atoms in the case of a single impurity), to explore the spatial extent of the YSR states. During the calculations in the superconducting state, we assumed an effective pair potential of $\Delta_{\rm eff}=1.51$ meV \cite{beck-natcomm2021} on the Nb atoms and $\Delta_{\rm eff}=0$ on the Mn impurities.

\subsection{Mn monomer on Nb(110)}\label{sec:monomer}

\begin{figure}[htb]
  \includegraphics[width=0.99\linewidth]{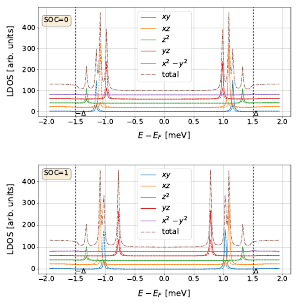}
\caption{The LDOS of the Mn adatom on the Nb(110) surface decomposed according to canonical $d$ orbitals without (top) and with (bottom) spin-orbit coupling. The solid lines correspond to electron densities and the dashed lines to hole densities. The total DOS summed for all orbitals and for the electrons and holes is plotted with dash-dotted line.  For a better visibility, the curves are shifted with respect to each other. }
\label{fig:mn1-dos}
\end{figure}

In the case of the Mn monomer, we considered a perpendicular direction of the magnetization and performed the calculations of the self-consistent normal state and of the LDOS in the superconducting gap as outlined above. Motivated by the simple tight-binding analysis \cite{beck-natcomm2021} we decomposed both the electron and hole LDOS according to orbitals related to the irreducible representations of the $C_{2v}$ point group. As the $s$ and $p$ orbitals provide negligible contributions to the LDOS in the SC gap, in Fig.~\ref{fig:mn1-dos} we only show the calculated $d$-like electron and hole contributions to the LDOS for the Mn impurity, both in the absence and in the presence of spin-orbit coupling. The former one is possible by manipulating the Dirac equation as proposed originally by Ebert {\em at al.} \cite{ebert}. In both cases, four pairs of sharp peaks in the LDOS are seen that clearly can be associated with  Yu-Shiba-Rusinov states induced by the magnetic impurity.
It is obvious  that the electron-hole symmetry of the DBdG equations is well reflected in our results. One can also observe that the states below the Fermi level are mostly hole-like, while the symmetric states with positive energy are mostly electron-like.

To get a deeper insight into the symmetry of the YSR states let us first consider the case without SOC. From the orbital resolution of the LDOS we conclude that three YSR states can be assigned with single $d$ orbitals: $d_{yz}$ at $\pm$0.98  meV, $d_{xz}$ at  $\pm$1.1  meV and $d_{xy}$ at $\pm$1.2 meV, while the peak at $\pm$1.3  meV having dominant $d_{z^2}$ character also has a small $d_{x^2-y^2}$ contribution. Since in the absence of SOC the Hamilton operator is diagonal in spin space, it is not surprising to recognize that the observed orbital decomposition of the YSR states reflects the (one-dimensional) irreducible representations of the simple $C_{2v}$ point group: the first three $d$ orbitals correspond to different irreducible representations, while both the  $d_{z^2}$ and the $d_{x^2-y^2}$ orbitals belong to the total symmetric $A_1$ representation, therefore, they can hybridize. According to this reasoning there should exist another   $d_{z^2}-d_{x^2-y^2}$ peak with $d_{x^2-y^2}$ dominance. Most possibly, this peak can not be seen as it might form outside the coherence region $(E_F-\Delta,E_F+\Delta)$, see later.

By switching on the SOC, the position of some of the peaks apparently changes: the $d_{yz}$ and the $d_{xy}$ peaks move closer to $E_F$ by about 0.2  meV  and 0.1 meV, respectively, while the $d_{xz}$ and $d_{z^2}-d_{x^2-y^2}$ peaks roughly keep their positions in energy. As a result, the order of the $d_{xy}$ and $d_{xz}$ peaks becomes reversed.
It is also expected that the SOC mixes both the spin and orbital characters of the YSR states. However, such mixing of the orbital characters of the YSR states is hardly visible in  the lower panel of Fig.~\ref{fig:mn1-dos}: only a minor peak in the $d_{xz}$ LDOS can be inferred at the position of the large $d_{yz}$ peak at $\pm0.75$ meV due to the spin-orbit coupling between states with the orbital quantum numbers $m_\ell=\pm 1$.
We repeated the DBdG simulations in the presence of SOC also with the effective potentials and fields obtained from the self-consistent calculations without SOC and found almost the same positions for the YSR states as without SOC in both steps of the calculations (upper panel of Fig.~\ref{fig:mn1-dos}).
From this observation we conclude that the shift of the YSR states due to the SOC (lower panel of Fig.~\ref{fig:mn1-dos}) originates from its indirect effect in changing the self-consistent potentials and fields.

As mentioned already, the magnetic impurity causes a spatial perturbation in the charge and magnetization density in the normal state, which manifests itself also  in the formation of local spin-polarization on the atoms around the impurity. Within the local density approximation (LDA) of density functional theory (DFT), this local spin-polarization is represented by the local exchange field being zero for a non-magnetic atom.
In order to investigate the role that the local spin-polarization in the nonmagnetic host plays in the formation of the YSR states we performed a series of calculations including an  increasing number of neighbor shells in the IC  with local exchange fields switched on.
In practice, in all calculations we used the same IC containing four neighbor shells and we  switching off the selfconsistently calculated exchange field at the same site. 

\begin{figure}[htb]
\includegraphics[width=0.99\linewidth]{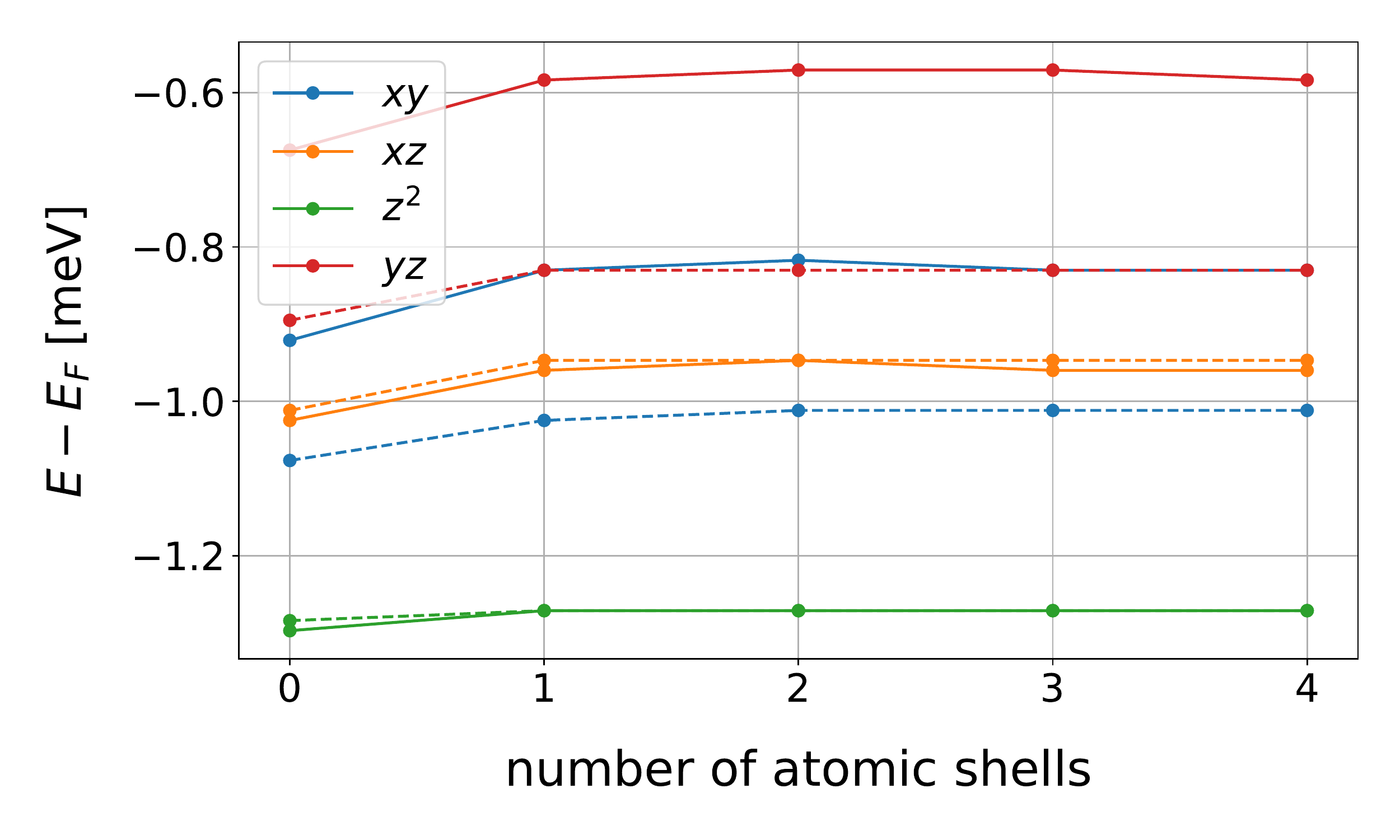}
\caption{ The energy of the YSR states below $E_F$ of the Mn adatom on Nb(110) as a function of the number of atomic shells around the adatom formed by host atoms with nonzero exchange field included in the calculations. The solid and dashed lines stand for the cases with and without spin-orbit coupling, respectively.}
\label{fig:roting}
\end{figure}

We start from a zero shell neighborhood meaning that at all sites in the IC but at the Mn impurity the exchange field is switched off and then we turn on the exchange fields determined from the normal-state self-consistent calculations shell by shell. The results are shown in Fig.~\ref{fig:roting}, where we plot the positions of the YSR peaks as a function of atomic shell number for which the exchange fields were included in the calculations. It can be seen that even in the zero shell calculation we get a rather accurate position of the YSR peaks, which changes only a little as we include the first shell. There is an even smaller change with the inclusion of the farther nearest-neighbor shells.
As can be seen from the figure, the above findings apply irrespective of whether the SOC is switched on or off, which is not surprising, since the formation of the local spin-polarization is usually insensitive of the SOC.

When comparing our results with the experimental findings in Ref. \onlinecite{beck-natcomm2021}, one can easily see that
regarding both the number of peaks and their positions they agree favorably.  The only peak which appears to be off in position is the one closest to the center of the gap, being shifted considerably towards the gap edge in our calculations. In order to determine the orbital character of the YSR states
the authors of Ref. \onlinecite{beck-natcomm2021} relied on the study of the spatial symmetry of the scanning tunneling spectroscopy (STS) pictures.
As discussed in context of Fig.~\ref{fig:mn1-dos},
in the current formalism it is straightforward to decompose the LDOS at the Mn impurity
according to orbitals corresponding to the irreducible representations of the point group of the system, in the case of Nb(110), of the $C_{2v}$ group.
The question then arises whether this decomposition can be correlated with the spatial shape of the STS pictures. To answer this question we
solved the DBdG equations for a large number of atoms in the first vacuum layer around the impurity site
and calculated the LDOS on each site in this IC. Due to the atomic sphere approximation used in our method,  a single averaged value of the LDOS can be attributed to each atomic cell.
Since such a visualization of the LDOS would be rather course grained, we applied some interpolation to the data within the atomic cells.

\begin{figure}[htb]
 \includegraphics[width=0.99\linewidth]{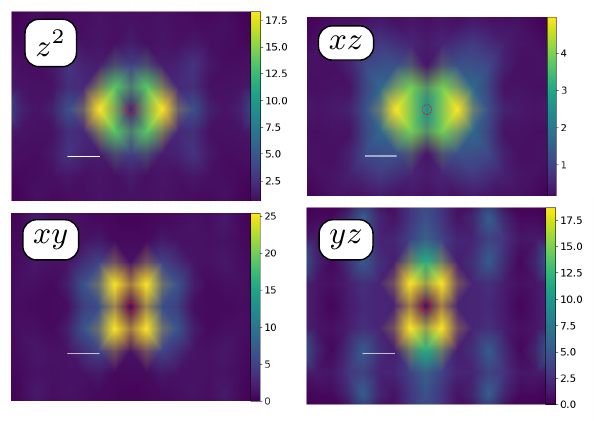}
\caption{The color map of the LDOS in arbitrary units in the vacuum layer containing the Mn adatom at the energies corresponding to the four YSR states identified in Fig.\ref{fig:mn1-dos}. The LDOS in a 2D unit cell is associated with the averaged LDOS from the corresponding atomic sphere. The LDOS of the Mn atom is replaced by the average of the LDOS from the first nearest neighbor vacuum cells. To get a smooth picture in space an interpolation scheme was used. The position of the Mn atom is marked by a small red circle. The white bar denotes a lattice spacing along the [100] direction ($a_{\rm 2D}$).
}
\label{fig:mn1-sts}
\end{figure}

We presented the obtained LDOS maps in Fig.~\ref{fig:mn1-sts}
at the energies of the YSR peaks seen in Fig. \ref{fig:mn1-dos} below the Fermi energy. These patterns reflect the spatial distribution, in particular, the symmetry of the corresponding YSR states. For example, the LDOS map of the $d_{xy}$ YSR peak at $-1.01$ meV displays zero values at the $y-z$ ($x=0$)  and the $x-z$ ($y=0$) planes, thus it can safely be associated with the $-\beta$ state of $d_{xy}$ character in Fig. 1. of Ref. \onlinecite{beck-natcomm2021}. Similarly, the $d_{yz}$ state at $-0.58$ meV, the $d_{xz}$ state at $-0.96$ meV and the $d_{z^2}$ state at $-1.28$ meV can be directly related to the $-\delta$, $-\gamma$ and $-\alpha$ states, respectively, detected experimentally and identified also in terms of tight-binding simulations\cite{beck-natcomm2021}.
Thus, our calculations reproduce all the YSR states of a Mn adatom on Nb(110) with the correct order in energy and with the correct orbital character seen from STS experiments\cite{beck-natcomm2021}.

As proposed originally by Shiba \cite{2}, the YSR states are known to vary in energy as a function of the atomic spin ($S$).
In order to study this relationship within a first principles multi-orbital scenario, we performed a computer experiment, where we artificially scaled the exchange field of the Mn atom and calculated the LDOS in the superconducting state as a function of the scaling factor, $B/B_0$. Because of the missing self-consistency, we simultaneously set the exchange fields to zero
for all the atoms in the extended impurity cluster. Note that a more sophisticated scaling could be obtained by the so-called "fixed moment search", where the size of the impurity's spin-moment is constrained, while the effective potentials and fields, as well as the induced exchange fields are recalculated self-consistently.

\begin{figure}[hbt!]
\includegraphics[width=.99\linewidth]{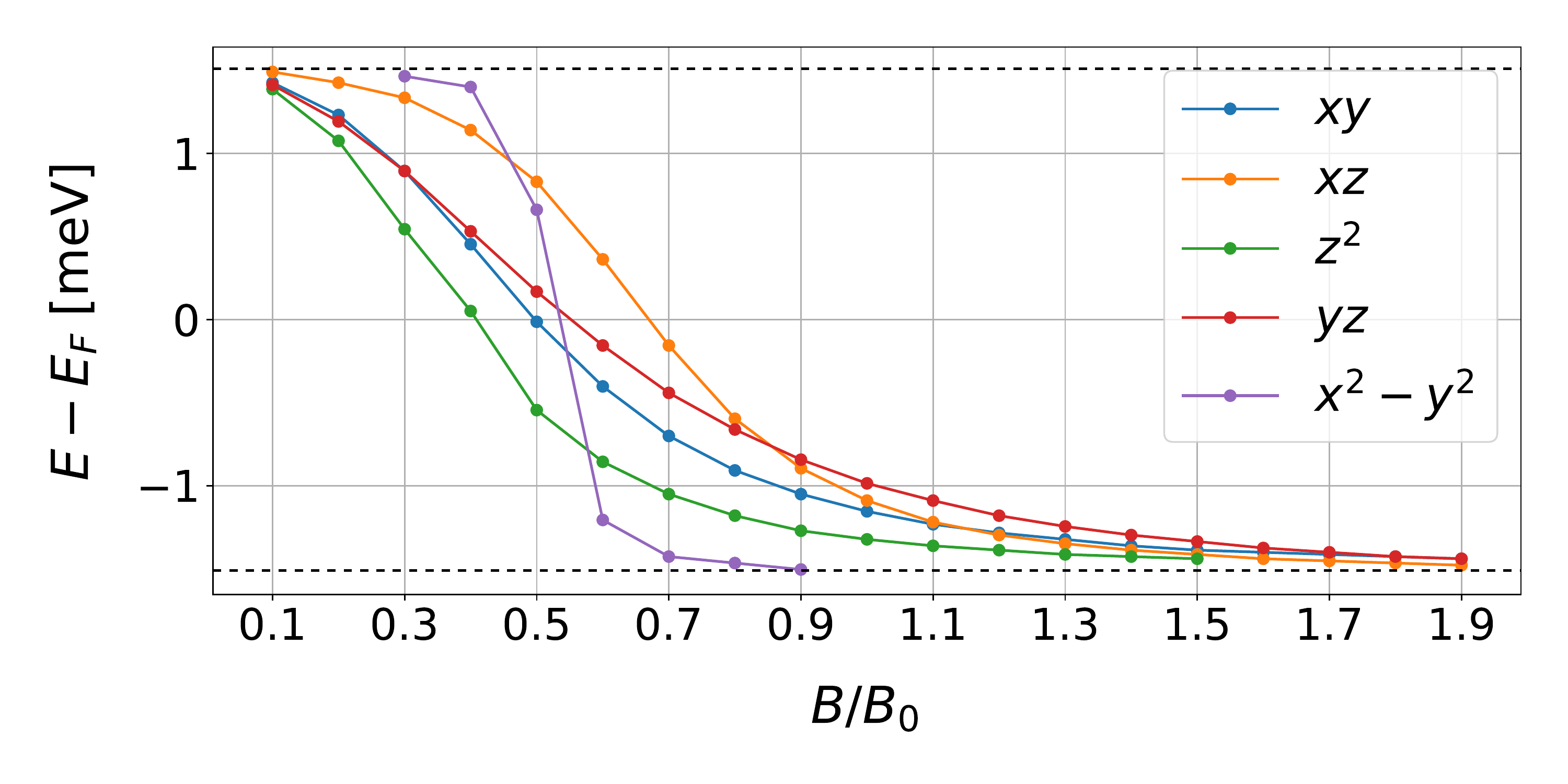}
    \caption{The energy of all the five $d$-like YSR states for the Mn adatom on Nb(110) as a function of the scaling of the exchange field, $B/B_0$, on the Mn atom. The calculations have been performed without SOC.
}
    \label{fig:bxc-scale}
\end{figure}

Fig.~\ref{fig:bxc-scale} shows the energy of the YSR states for $0<B/B_0<2$  using the simplified scaling picture described above.  One can see that the peaks enter into the energy range of the superconducting gap at rather low fields and as we increase the exchange field they cross the Fermi energy and then leave the gap. This trend is compatible with the original prediction of Shiba, even though a direct comparison is quite difficult. One important feature of the figure that the scattering chanel with $x^2-y^2$ orbital character appears only in a rather narrow range of the exchange field, and at the actual self-consistent value ($B/B_0=1.0$) it does not provide any contribution to the density of states within the gap. This is in perfect agreement with the experimental findings of Beck  et al. \cite{beck-natcomm2021}, who could not find a peak with $x^2-y^2$ character.

\subsection{Mn dimers on Nb(110)}\label{sec:dimer}
In the case of two impurities forming a dimer, three adjacent positions were reported in Ref. \onlinecite{beck-natcomm2021} that can be characterized by the direction of the vectors connecting the two Mn atoms, namely, the nearest-neighbor (NN) position by [100], the second NN neighbor position by [1$\bar{1}0$], and the third nearest NN position by [1$\bar{1}$1].  Here we consider the  [1$\bar{1}0$] the [1$\bar{1}$1] dimers only.
Based both on experiments on short Mn chains on Nb(110)\cite{schneider-2021} and on ab-initio calculations \cite{laszloffy-2021} it is suggested that the coupling of the Mn atoms in the [1$\bar{1}$0] dimer is ferromagnetic (FM), while in the [1$\bar{1}1$] dimer it is antiferromagnetic (AFM).

\begin{figure}[htb]
\includegraphics[width=0.99\linewidth]{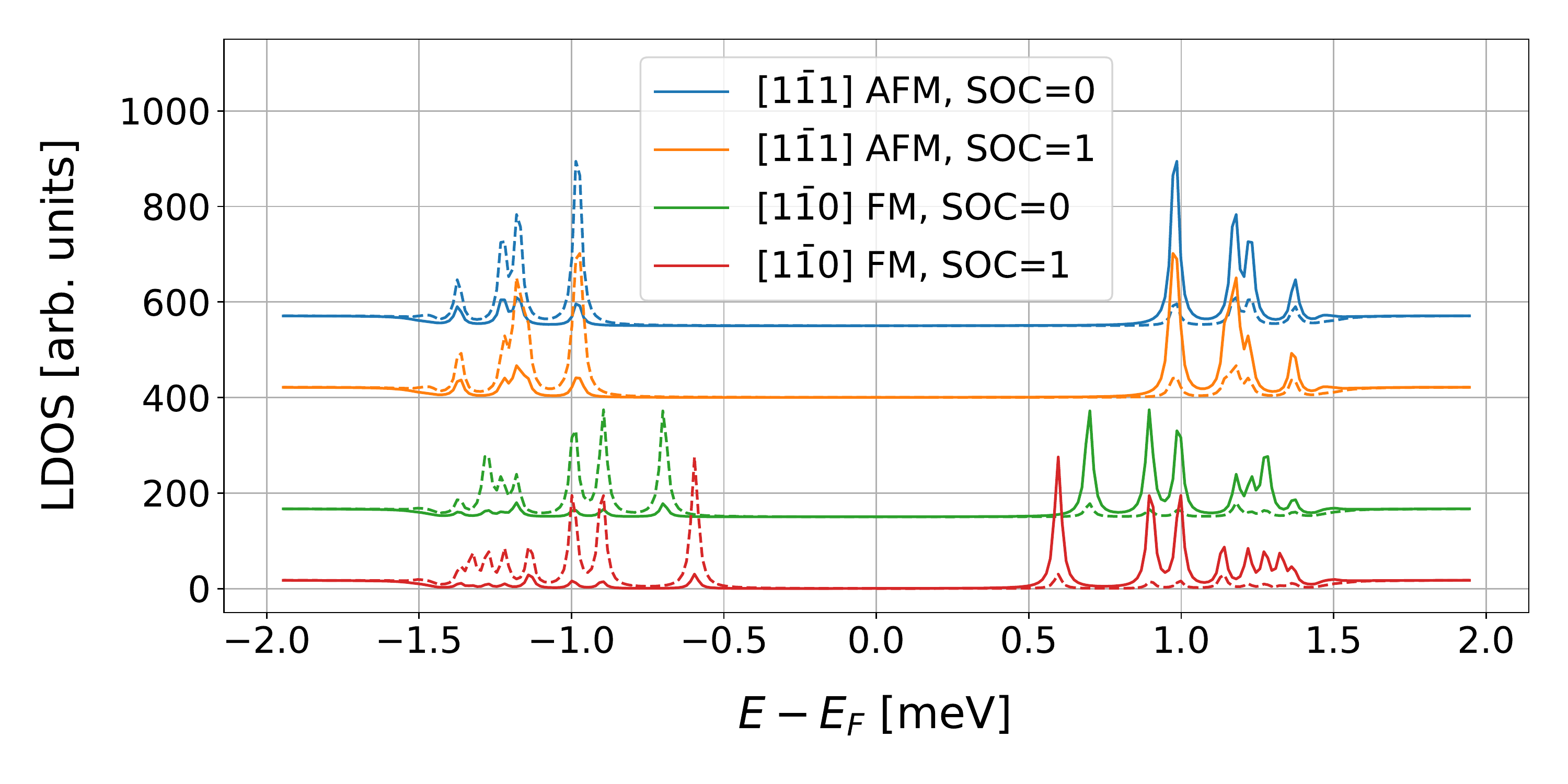}
\caption{LDOS in the superconducting state calculated without and with SOC for the ferromagnetic [1$\bar{1}$0] and the antiferromagnetic [1$\bar{1}$1] Mn dimers on Nb(110). Solid and dashed lines stand for the electron and hole part of the LDOS.
}
\label{fig:dimer-xd}
\end{figure}

The LDOS for these two dimers is plotted in Fig.~\ref{fig:dimer-xd}  both without and with SOC taken into account in the calculations. Let's consider first the results without SOC.
The LDOS for the FM [1$\bar{1}$0] dimer shows a complex structure formed by eight peaks each for negative and positive energies upon electron-hole symmetry, which is twice the number of the peaks as seen in the case of a single impurity.
The number of these states and their positions are again in reasonable agreement with experiment\cite{beck-natcomm2021}, though - similarly to the single impurity case - the peaks appear somewhat shifted towards the gap edges.
On the contrary, in the case of the AFM [1$\bar{1}$1] dimer, the YSR states appear in a very similar way as for a single impurity density of states on Fig.~\ref{fig:mn1-dos} and the doubling of the number of YSR states can not be observed.  This is in contrast with the experiment where six peaks were detected\cite{beck-natcomm2021}.

As well-known, if the localized YSR states of the two impurities overlap, they can hybridize leading to the formation of symmetric or bonding and anti-symmetric or anti-bonding pairs, where the symmetric/antisymmetric one has an enhanced/zero density at the mirror plane between the two impurities.
However, at least within the a non-relativistic theory\cite{3}, such a hybridization and splitting of the YSR states occurs only in FM-coupled dimers, while the YSR states remain degenerate in an AFM-coupled dimer as reported also experimentally \cite{flatte,morr}.

\begin{figure}[h!]
    \centering
     \includegraphics[width=.95\linewidth]{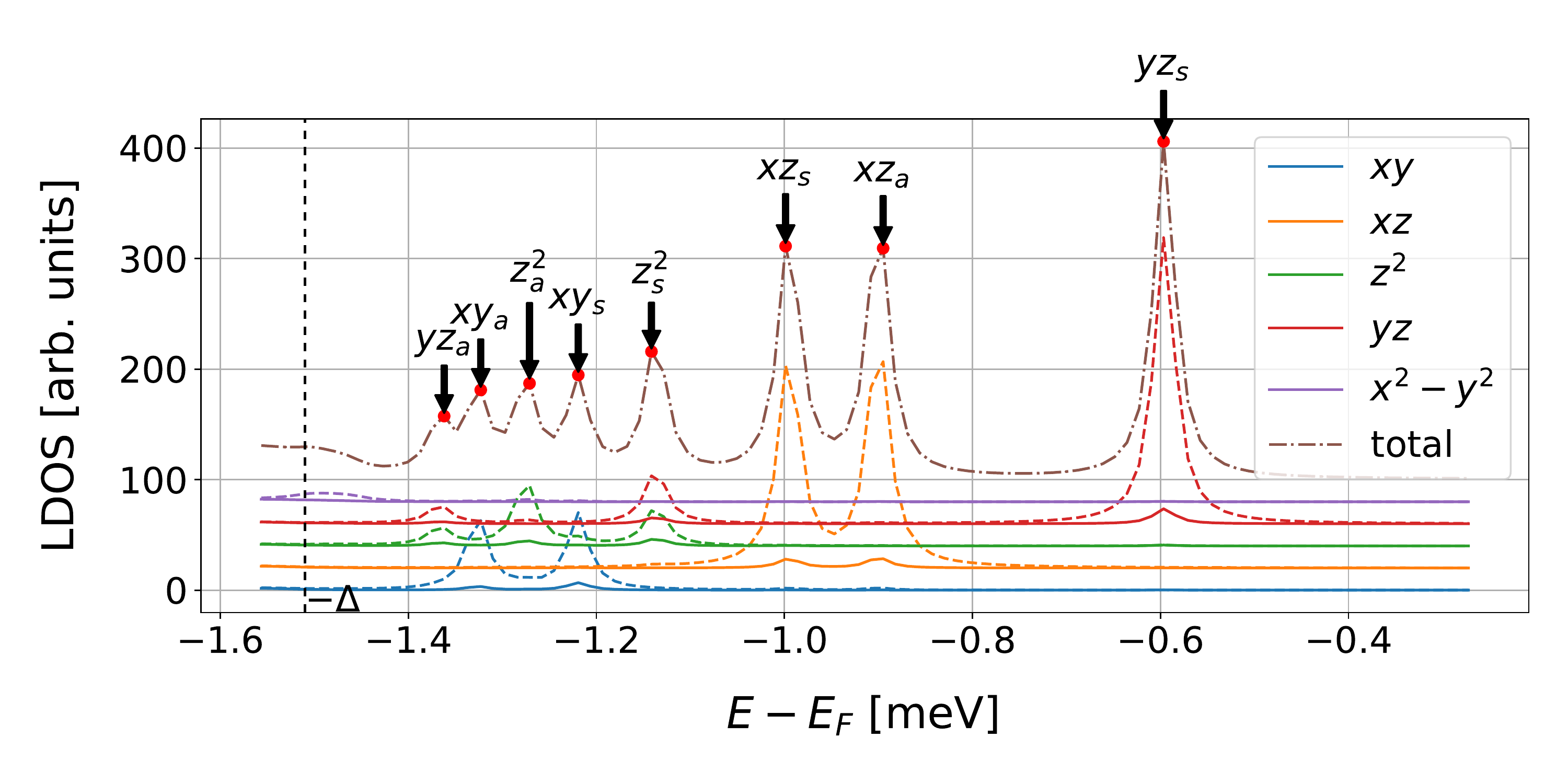} \quad \,
        \includegraphics[width=.99\linewidth]{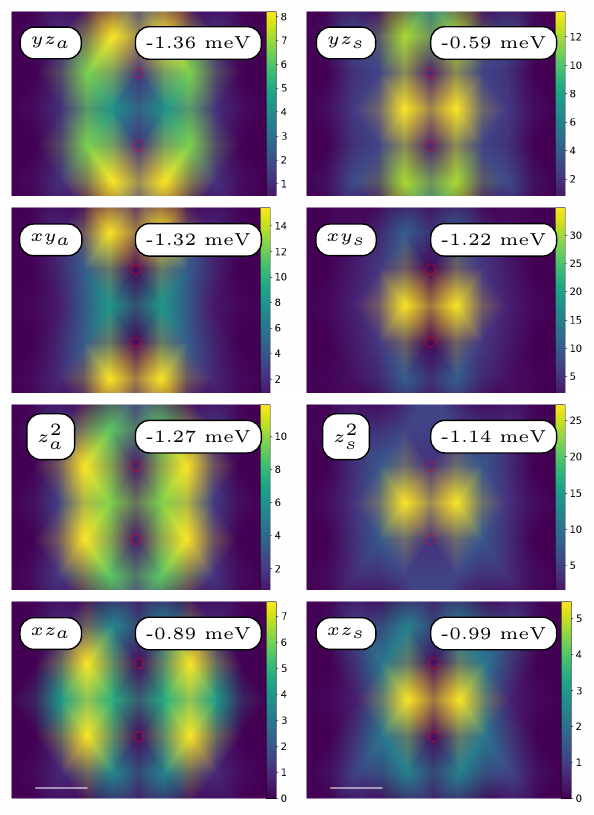}
    \caption{Upper panel: The decomposition of the LDOS below $E_F$ to $d$-like orbitals for the FM  [1$\bar{1}$0] dimer. The classification of the split YSR peaks is also displayed as obtained from the simulated STS pictures.
    Lower panels:  Color map of the simulated STS pictures (see Fig. 3) corresponding to the eight YSR peaks marked in the upper panel.  The left/right four entries refer to the antisymmetric/symmetric YSR states of the dimer. The positions of the Mn atoms are marked by small red circles. The white bar denotes a lattice spacing along the [100] direction ($a_{\rm 2D}$).
    }
    \label{fig:mn2x-dos-sts}
\end{figure}

To analyze the hybridizations and the splitting of the YSR states for the FM [1$\bar{1}$0] dimer in more details, in the top panel of Fig.  \ref{fig:mn2x-dos-sts} we plot the decomposition of the LDOS into $d$-like contributions near the lower edge of the SC gap.
In addition, we plotted the color maps of the LDOS in the vacuum layer containing the Mn dimer, see also  Fig. 3 for the single impurity. Based on these figures, the peaks at -1.22 meV and -1.32 meV with large $xy$ contributions could easily be identified as the  symmetric and antisymmetric components of the split $xy$ YSR state, denoted by $xy_s$ and $xy_a$, respectively. Similarly, the $xz_a$ and $xz_s$ states can be straightforwardly recognized because of their overwhelming $xy$ contributions and the shapes in the lowest  two panels in  Fig.  \ref{fig:mn2x-dos-sts}. The peak at -0.59 meV has dominant $yz$ character and from the corresponding entry in  Fig.  \ref{fig:mn2x-dos-sts} it can be identified as the bonding $yz_s$ state. At -1.27 meV one can see a peak with dominant $z^2$ contribution, somewhat hybridized with $yz$ states. Nevertheless, based on the respective color map in Fig.  \ref{fig:mn2x-dos-sts} we can safely associate it with the antisymmetric $z^2_a$ state. There are two peaks remaining at -1.36 meV and -1.14 meV, both of them containing  almost equal $yz$ and $z^2$ orbital contributions. From the corresponding color maps it is clear that they display an antisymmetric and a symmetric hybridized state, referred to as the $yz_a$ and $z_s^2$ states, respectively.

The above analysis demonstrates that it is possible to classify the split multi-orbital YSR states of the FM dimer derived from our first principles theory and they show similar patterns
as the experiments\cite{beck-natcomm2021}. In contrast to the experiments, in our calculations the symmetric states are higher in energy than the corresponding antisymmetric ones. However, note that for their positive-energy counterparts containing mostly electron states, this relationship is reserved. It is also worthwhile to mention that the splitting of the $xy$ ($\beta$) states could not be resolved in the experiment.

Inferring the LDOS's in Fig. \ref{fig:dimer-xd} with SOC turned on, it is obvious that
 for both kind of dimers the SOC only slightly alters the positions of the YSR states. In case of the FM coupled [1$\bar{1}$0] dimer it seems that additional splittings happen due to the SOC, but a more detailed investigation shows that
 this feature happens due to non-uniform shifts of the peak positions with different orbital characters. In case of the AFM [1$\bar{1}$1] dimer the changes caused by the SOC are hardly visible. In sharp contrast with the theoretical explanation  in Ref. \onlinecite{beck-natcomm2021}, we conclude that the spin-orbit coupling in this system is not strong enough to explain the splitting of the YSR states found experimentally for the AFM dimer.
 This conclusion is consistent with our previous findings for the Mn adatom, where we also found a rather weak effect of SOC, and explains the similarity between the LDOS of the single impurity and the AFM coupled dimer as it was put forward earlier in the context of a non-relativistic theory\cite{3}.

\begin{figure}[htb]
\includegraphics[width=0.99\linewidth]{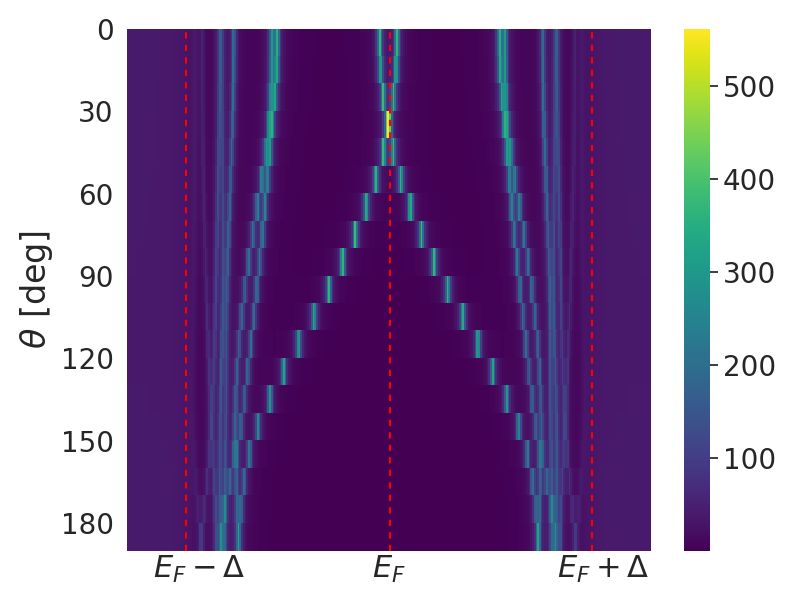}
\caption{The energy of the YSR states in the [1$\bar{1}$1] Mn dimer on Nb(110) as a function of the relative angle $\theta$ between the two Mn moments. Note that the FM and the AFM states correspond to $\theta=0$ and $\theta=180^\circ$, respectively.}
\label{fig:mn2d-rot}
\end{figure}

The theoretical study of Lászlóffy {\em et al.}\cite{laszloffy-2021} suggests that a slightly non-collinear and/or canted configuration can exist for the dimers in the normal state.
The possibility of the formation of such tilted ground states and the lack of split YSR states in the AFM dimer compared to experiments motivated us to study the effect of the non-collinear spin-configurations to the YSR states. In these calculations we kept fixed the effective potentials and fields obtained selfconsistently in the collinear state and rotated the direction of one of the Mn moments continuously between the AFM ($\theta=180^\circ$) and the FM ($\theta=0^\circ$) states. In Fig. \ref{fig:mn2d-rot} we show the change of the energy of the YSR states for the [1$\bar{1}$1] dimer as a function of the opening angle $\theta$.

As also shown based on tight-binding calculations in Ref. \onlinecite{beck-natcomm2021}, a deviation from the collinear magnetic configuration lifts the degeneracy of the YSR states present in the collinear AFM dimer.
 In Table \ref{table:d-dimer} we presented the splittings calculated for the relative angles $170^\circ$ and $160^\circ$, together with the those found experimentally\cite{beck-natcomm2021}. Apparently, for these angles the calculated values are in the range of the measured ones, with exception of the splitting of the $\gamma$ state.
 Nevertheless, such big deviations from the AFM spin configuration seem to be unrealistic based on the spin-model simulations of Ref. \onlinecite{laszloffy-2021}, where for the NN [1$\bar{1}$1] dimer a $\theta \simeq 179.5^\circ$ was predicted.

Another important feature that can be seen in Fig. \ref{fig:mn2d-rot} that one line of the split $d_{yz}$ $(\delta)$ states moves rapidly towards the Fermi level, while the energies of the other YSR states are shifted more moderately when going from the AFM to the FM configuration.
At roughly $\theta=40^\circ$ this peak can be found right in the middle of the energy range, namely at zero bias. Such a behavior of the upper lying $yz$ peak has been also found in Ref. \onlinecite{beck-natcomm2021} using a multi-orbital tight-binding approach, though, because of the much deeper position of the  $\delta$-states, already at about $\theta=140^\circ-150^\circ$ depending on whether the SOC was turned off or on.
Zero bias peaks (ZBP) have the property that they are necessarily have an equal
electron-like and hole-like component in the LDOS. Such zero bias peaks in extended chains are of primary interest nowadays because they are supposed to realize topological quantum bits\cite{beenakker-2011,alicea-2012,elliott-2015}. Quite interestingly, theoretical calculations on the same platform as in this work reveal that ZBP's can be formed in single magnetic impurities on top of superconducting Pb(110) by tilting the magnetic orientation of the impurity with respect to the surface \cite{kyungwha-2021}.  The present calculations demonstrate that ZBP's can be created with the manipulation of the relative angle in dimers. It is therefore expected that non-collinearity, e.g. spin-spiral states, can induce topological ZBP's in long chains.

\begin{table}[htb]
    \caption{Calculated energy splittings of the YSR states in the [1$\bar{1}$1] Mn dimer on Nb(110) for Mn moments with a relative angle $\theta$. The last line contains the experimental values from Ref. \onlinecite{beck-natcomm2021}. }
    \label{table:d-dimer}
    \begin{ruledtabular}
        \begin{tabular}{c|cccc}
         $\theta$ & $z^2$ ($\alpha$) & $xy$ ($\beta$) & $xz$ ($\gamma$) & $yz$ ($\delta$) \\ \hline
         170$^\circ$ & 37  & 70 & 41 & 19\\
         160$^\circ$ & 84  & 40 & 39  & 53\\ \hline
         exp & 130 & 400 & - & 110\\
        \end{tabular}
    \end{ruledtabular}
\end{table}

\section{Summary and conclusions}
In order to provide a material specific description of magnetic impurities on superconducting surfaces, in the present paper we generalised the Embedding Method of Multiple Scattering Theory to solve the Dirac-Bogoliubov-de Gennes equations. By using  the new method, first-principles band structure, relativistic effects and superconductivity is treated on the same footing. In particular, the multi-orbital treatment of the in-gap states is {\em a priori} inherently involved in the theory.

For a single Mn impurity placed on Nb(110) we found a series of Yu-Shiba-Rusinov states in the superconducting gap, in excellent agreement with experiments both concerning the energy and the orbital decomposition of these states. Moreover, we were able to prove  that the orbital character of the YSR states inferred from the LDOS of the Mn impurity is directly related to the spatial shape and symmetry of the corresponding simulated STS pictures. Our calculations demonstrate that the spin-orbit coupling has quite a little effect in shifting the energy of these states. We also showed that the local spin-polarization on the magnetic impurity beyond the first coordination shell influence negligibly the energy of the YSR states.  In line with earlier theoretical predictions, we also demonstrated that the positions of the YSR states are very sensitive on the size of the magnetic moments, which can be used to explain the missing in-gap state with $d_{x^2-y^2}$ orbital character in case of the Mn adatom on Nb(110).

In case of the close packed [1$\bar{1}$0] ferromagnetic Mn dimer, our calculations demonstrated the well-established splitting of the YSR states due to hybridization.
By analyzing the symmetry of the simulated STS maps we managed to identify the orbital character of the split states and found a satisfactory agreement with the experiment. Quite interestingly, however, we obtained a reversed order for the energies of the bonding and antibonding states as compared to the experiment.

In good agreement with the prediction of non-relativistic theories we found no splitting of the YSR states for the antiferromagnetic [1$\bar{1}$1] Mn dimer when the SOC was switched off in the calculations.
By including the SOC, we observed very small changes in the energy of the YSR states, in sharp contradiction with the experiment and related tight-binding calculations including SOC which showed a sizeable splitting of the YSR states also in this case. Our first principles calculations thus indicate that the SOC is not sufficiently strong to explain every detail of the experimental findings for the AFM coupled [1$\bar{1}$1] Mn dimer.

As a possibility to resolve this contradiction, we investigated the effect of the relative magnetic orientation in the dimer and found splittings of the YSR states being comparable with the measured ones for about $20^\circ$ deviation from the AFM arrangement. Since such spin-configurations seem to be unrealistic to form the ground state of the Mn dimer, further investigations are needed to improve the agreement between theory and experiment.
 It is a subject of future work to take into account e.g. the relaxation of the lateral position of the impurities, where an increased hybridization with the substrate is expected and a more accurate description of the dimer YSR states can be obtained.

Remarkably, at about $40^\circ$ to the parallel alignment of the moments the energy of the deepest YSR state reached the Fermi level, forming thus a zero bias peak. This hints the possibility of creating zero bias peaks in longer chains upon non-collinear arrangements of the spin.


\begin{acknowledgments}
Among the authors B.Ny., A.L., L.Sz. and B.U. are grateful for the financial support by the Ministry of Innovation and Technology and the Hungarian National Scientific Research Fund under project No.\ K131938 and within the Quantum Information National Laboratory of Hungary. B.Ny. also greatful for the support of ÚNKP-20-3-II-BME-249 New National Excellence Program of the Ministry for Innovation and Technology of Hungary. G.Cs. gratefully acknowledges support from the European Union’s Horizon 2020 research and innovation program under the Marie Sklodowska-Curie Grant Agreement No. 754510. This work was supported by Spanish MINECO (the Severo Ochoa Centers of Excellence Program under Grant No. SEV- 2017-0706), Spanish MICIU, AEI and EU FEDER (Grant No. PGC2018-096955-B-C43), and Generalitat de Catalunya (Grant No. 2017SGR1506 and the CERCA Program). The work was also supported by the European Union MaX Center of Excellence (EU-H2020 Grant No. 824143).
\end{acknowledgments}


\begin{thebibliography}{99}

\bibitem{1} L. Yu, Acta Phys. Sin. \textbf{21}, 75 (1965).
\bibitem{2} H. Shiba, Prog. Theor. Phys. \textbf{40}, 435 (1968).
\bibitem{3} A. I. Rusinov, Zh. Eksp. Teor. Fiz. Pisma Red. 9, 146 (1968) [JETP Lett. 9, 85 (1969)].
\bibitem{4} A. L. Fetter, Phys. Rev. \textbf{140}, A1921 (1965).
\bibitem{5} A. Yazdani, B. A. Jones, C. P. Lutz, M. F. Crommie, and D. M. Eigler, Science \textbf{275}, 1767 (1997).
\bibitem{6} A. V. Balatsky, I. Vekhter, and J.-X. Zhu, Rev. Mod. Phys. \textbf{78}, 373 (2006).
\bibitem{9} S.-H. Ji, T. Zhang, Y.-S. Fu, X. Chen, X.-C. Ma, J. Li,W.-H. Duan, J.-F. Jia, and Q.-K. Xue, Phys. Rev. Lett. \textbf{100}, 226801 (2008).
\bibitem{10} M. Ruby, F. Pientka, Y. Peng, F. von Oppen, B.W. Heinrich and K. J. Franke, Phys. Rev. Lett. \textbf{115}, 087001 (2015).
\bibitem{11} N. Hatter, B.W. Heinrich, M. Ruby, J. I. Pascual, and K. J. Franke, Nat. Commun. \textbf{6}, 8988 (2015).
\bibitem{100} G. Csire, B. Újfalussy, J. Cserti, and B. L. Gyorffy.  Phys. Rev. B, \textbf{91}, 165142, (2015).
\bibitem{101} G Csire, A. Deák, B. Nyári, H. Ebert, J. Annett, and B. Ujfalussy. Phys. Rev. B, \textbf{97}, 024514 (2018).
\bibitem{lazarovits-2002} B. Lazarovits, L. Szunyogh, and P. Weinberger.  Phys. Rev. B, \textbf{65}, 104441, (2002).
\bibitem{csire-relativistic2018} G. Csire, B. Ujfalussy, and J. F. Annett.  The European Physical Journal B, 91(10),217, (2018).
\bibitem{104} G. Csire, J. Cserti, I Tütto and B. Ujfalussy.  Phys.Rev. B, 94:104511, 2016.
\bibitem{105} Gábor Csire, Stephan Schönecker, and Balázs Újfalussy. First-principles approach to thin
superconducting slabs and heterostructures. Phys. Rev. B, 94:140502, Oct 2016.
\bibitem{106} S. K. Ghosh, G. Csire, P. Whittlesea, J. F. Annett, M. Gradhand,
B. Ujfalussy and J. Quintanilla.  Phys. Rev. B, 101, 100506, (2020).
\bibitem{107} T. G. Saunderson, J. F. Annett, B. Újfalussy, G. Csire, and M. Gradhand,
Phys. Rev. B {\bf 101}, 064510, 2020.
\bibitem{saunderson-2021} T. G. Saunderson, J. F. Annett, G. Csire, and M. Gradhand, arXiv:2107.00237 (2021).
\bibitem{vosko-1980} S.~H.~Vosko, L.~Wilk, and M.~Nusair, Can. J. Phys. \textbf{58}, 1200 (1980).
\bibitem{straumanis-1970} M.~E.~Straumanis, and S.~Zyszczynski, J. Appl. Crystallogr., \textbf{3}, 1 (1970)
\bibitem{beck-natcomm2021} P.~Beck, L.~Schneider, L.~Rózsa, K.~Palotás, A.~Lászlóffy, L.~Szunyogh and R.~Wiesendanger, Nature Communications, \textbf{12}(1), 1-9 (2021).
\bibitem{ebert} H. Ebert and H. Freyer and A. Vernes and G.-Y. Guo, Phys. Rev. B, {\bf 53} 7721 (1996)
\bibitem{schneider-2021} Schneider, L., Beck, P., Wiebe, J. and Wiesendanger, R. Sci. Adv. 7, eabd7302 (2021).
\bibitem{laszloffy-2021}
A. Lászlóffy, K. Palotás, L. Rózsa, and L. Szunyogh,
Nanomaterials {\bf 11}, 1933 (2021).
\bibitem{flatte}
M. E. Flatté  and D. E. Reynolds, Phys. Rev. B {\bf 61}, 14810–14814 (2000).
\bibitem{morr} D. K. Morr and N. A. Stavropoulos,  Phys. Rev. B {\bf 67} 020502 (2003).
\bibitem{beenakker-2011}C. W. J. Beenakker, Annu. Rev. Condens. Matter Phys. {\bf 4}, 113–136 (2011).
\bibitem{alicea-2012} J. Alicea, Reports Prog. Phys. {\bf 75}, 76501 (2012).
\bibitem{elliott-2015} S. R. Elliott and M. Franz, Rev. Mod. Phys. {\bf 87}, 137–163 (2015).
\bibitem{kyungwha-2021} K. Park, B. Nyári, A. Lászlóffy, L. Szunyogh, and B. Újfalussy, submitted for publication.
\end{thebibliography}
\end{document}